\documentclass[11pt]{article}

\def\lesssim{\mathrel{\hbox{\rlap{\hbox{\lower4pt\hbox{$\sim$}}}\hbox{$<$}}}}

\input{psfig.sty}
\usepackage{times}
\usepackage{cospar}
\begin{document}

\title{THE DYNAMIC BEHAVIOR OF SOFT GAMMA REPEATERS}


\author{Peter~M.~Woods}

\maketitle

\begin{center}
{\it Universities Space Research Association \\ 
National Space Science and Technology Center \\ 320 Sparkman Dr. 
Huntsville, AL 35805}
\end{center}

\begin{abstract}

Soft Gamma Repeaters (SGRs) undergo changes in their pulse properties and
persistent emission during episodes of intense burst activity.  Both
SGR~1900$+$14 and SGR~1806$-$20 have shown significant changes in their
spin-down rates during the last several years, yet the bulk of this variability
is not correlated with burst activity.  SGR~1900$+$14 has undergone large
changes in flux and a dramatic pulse profile change following burst activity in
1998.  The flux level of SGR~1627$-$41 has been decreasing since its only
recorded burst activity.  Here, we review the global properties of SGRs as well
as the observed dynamics of the pulsed and persistent emission properties of
SGR~1900$+$14, SGR~1806$-$20 and SGR~1627$-$41 during and following burst
active episodes and discuss what implications these results have for the burst
emission mechanism, the magnetic field dynamics of magnetars, the nature of the
torque variability, and SGRs in general.

\end{abstract}

\section{ INTRODUCTION}

Soft Gamma Repeaters (SGRs) are a small class (4 confirmed and one candidate)
of high-energy transient discovered through their emission of bright
X-ray/$\gamma$-ray bursts which repeat on timescales of seconds to years. 
Originally confused with Gamma-Ray Bursts (GRBs see e.g.\ Fishman and Meegan
1995), SGRs get their name from their burst properties that distinguish them
from classical GRBs; namely their softer spectral energy distribution and
repetition of bursts. 

In addition to being prolific burst emitters, SGRs are also persistent X-ray
sources.  For three objects, the X-ray radiation is modulated by the spin of
the underlying neutron star.  In each case, the star is spinning down rapidly,
indicative of a strong magnetic field (Kouveliotou et al.\ 1998a).  In fact, it
was this observation that clinched the now widely accepted magnetar model for
SGRs (Thompson and Duncan 1995, 1996).  A magnetar is a strongly magnetized
neutron star ($B_{\rm dip}$ $\sim 10^{14} - 10^{15}$ G) where the magnetic
energy exceeds all other sources of free energy in the system.  Unlike ordinary
radio pulsars, the rotational energy loss in SGRs is insignificant compared to
their overall energy output.  It is the decay of the strong field of the SGR
that powers both their burst and persistent emissions.

In addition to the four confirmed SGRs, there is another small class of neutron
stars dubbed the Anomalous X-ray Pulsars (AXPs) that were proposed by Thompson
and Duncan (1996) as magnetar candidates.  The AXPs share a number of
characteristics with the SGRs in terms of their persistent X-ray emission
properties, and two members were recently discovered to emit SGR-like bursts
(Gavriil et al.\ 2002; Kaspi et al.\ 2002).  The recent burst detection results
solidify the common nature of AXPs and SGRs.  A review of AXPs will be
presented within this volume (Gavriil et al.\ 2003).

Here, I will review the salient properties of SGRs including burst
characteristics, persistent and pulsed X-ray emission, and the effects of burst
activity on SGR persistent emission properties.  Associations with supernova
remnants and massive star clusters will be discussed elsewhere in this volume
(Gaensler 2003).

\section{ BURST PROPERTIES}

The defining behavior of SGRs is their repetitive emission of luminous soft
$\gamma$-ray bursts.  Typically, the bursts last $\sim$0.1 sec and have energy
spectra (E $>$25 keV) that can be modeled with an Optically Thin Thermal
Bremsstrahlung (OTTB) having temperatures 20$-$40 keV (e.g.\
{G\"o\u{g}\"u\c{s}} et al.\ 2001, Aptekar et al.\ 2001).  At lower energies,
however, this model fails to fit the spectrum (e.g.\ Olive et al.\ 2003).  The
spectral temperatures show little or no variation with intensity, between
bursts, at different epochs, or between sources.  The burst energies follow a
power-law number distribution (dN/dE $\propto$ E$^{-5/3}$ [Cheng et al.\ 1996,
{G\"o\u{g}\"u\c{s}} et al.\ 1999) from $\sim$10$^{35}$ ergs up to at least
$\sim$10$^{42}$ ergs, consistent with a so-called self-organized critical
system (e.g.\ earthquakes, Solar flares, etc.) where the burst energy reservoir
greatly exceeds the energy output within any given burst.

Burst active states of SGRs, often referred to as outbursts, range in duration
from a few days to several months.  During these episodes, they emit anywhere
from a handful to several hundred bursts.  The burst rate histories of the four
SGRs spanning three decades are shown in Figure 1.  These lightcurves are
comprised of burst detections from several different detectors having different
sensitivity limits.  The non-uniformity of the burst sampling becomes
significant when comparing separate active episodes that were viewed with
different detectors.  For example, given a stationary burst energy (and peak
flux) distribution with a power-law slope of $-5/3$, an increase in the
detector sensitivity by a factor 10 yields a factor $\sim$5 increase in the
overall rate.

\vspace{-0.35in}

\begin{figure}[!htb]

\centerline{
\psfig{file=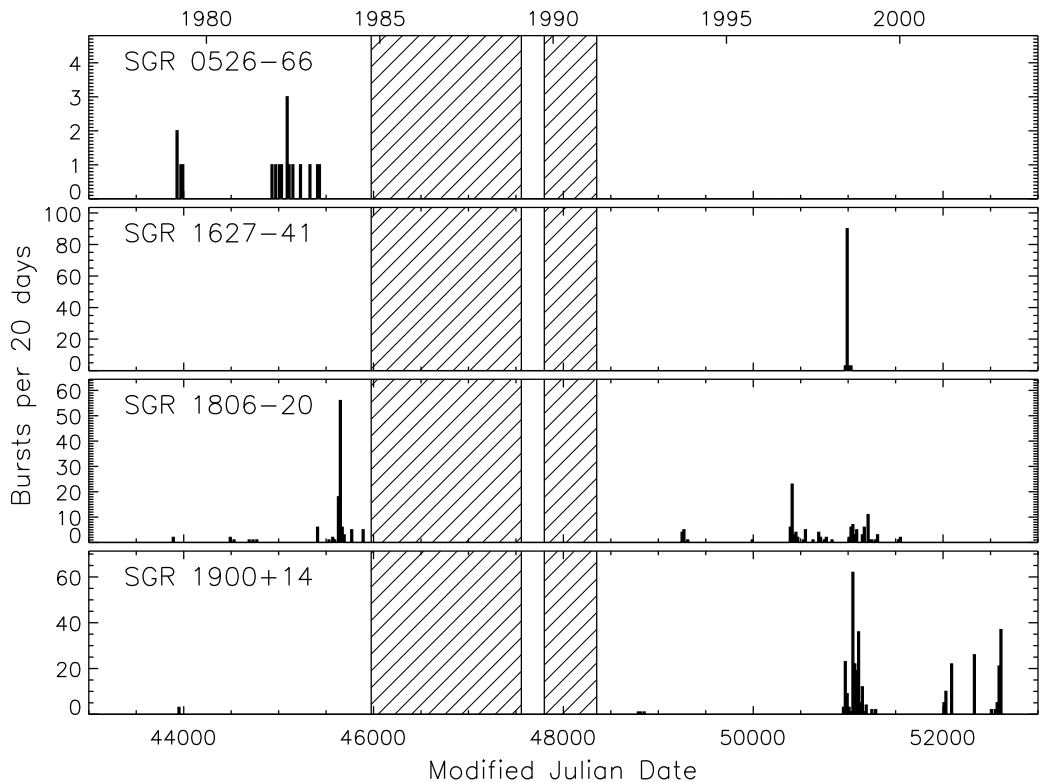,height=4.5in,angle=90}}
\vspace{-0.3in}

\caption{Burst activity lightcurve of the four confirmed SGRs. 
The bursts identified here were detected with various detectors having
different sensitivities.  The shaded regions indicate epochs where there were
no active detectors in flight to detect SGR bursts.  IPN data courtesy of Kevin
Hurley.}

\end{figure}

On two occasions, more energetic bursts or giant flares were recorded one each
from SGR~0526$-$66 on 1979 March 5 (Mazets et al.\ 1979) and SGR~1900$+$14 on
1998 August 27 (Hurley et al.\ 1999a).  The lightcurve of the August 27 flare
is shown in Figure 2.  Each of these extraordinary events had a bright
($\sim$10$^{44}$ ergs s$^{-1}$), spectrally hard initial spike followed by a
softer, several minute long tail showing coherent pulsations at 8 and 5 s,
respectively.  More recently, an intermediate flare ($\sim$10$^{43}$ ergs)
lasting 40 s was recorded from SGR~1900$+$14 on 2001 April 18 (Guidorzi et al.\
2001), suggesting a continuum of bursts energies rather than a dichotomy of
bursts and flares (Kouveliotou et al.\ 2001).

\vspace{-0.3in}

\begin{figure}[!htb]

\centerline{
\psfig{file=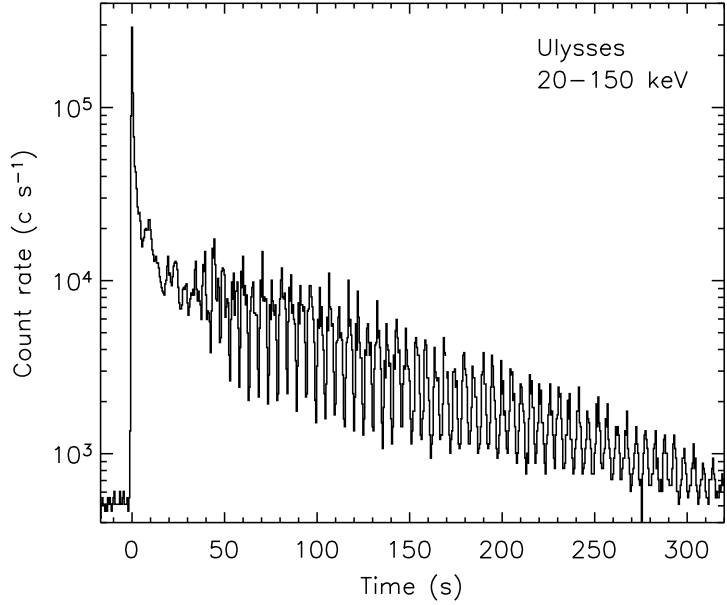,height=3.4in}}
\vspace{-0.1in}

\caption{The giant flare from SGR~1900$+$14 as observed with the gamma-ray
detector aboard Ulysses (20$-$150 keV).  Note the strong 5.16 sec pulsations
clearly visible during the decay.  Data courtesy of Kevin Hurley.}

\end{figure}

Recently, discrete features in the spectra of bursts from two SGRs, 1900$+$14
and 1806$-$20 have been reported.  In the case of SGR~1900$+$14, an emission
line at $\sim$6.5 keV was discovered (Strohmayer and Ibrahim 2000) during the
onset of a single, bright burst.  No candidate line features in other bursts
have been reported from this source since.  For SGR~1806$-$20, an absorption
line feature was found near $\sim$5 keV in a small subset of spectra taken from
selected bursts (Ibrahim et al.\ 2002).  The presence of less significant
features at higher energies consistent with a harmonic relationship is
intriguing, but their significance (for individual bursts) is not convincing.

\section{ PERSISTENT EMISSION}

All SGRs are associated with persistent X-ray counterparts; three of them have
quiescent luminosities $\sim$10$^{35}$ ergs s$^{-1}$ (Murakami et al.\ 1994;
Rothschild et al.\ 1993; Hurley et al.\ 1999b), while the quiescent flux level
of SGR~1627$-$41 has not yet been reached (C.\ Kouveliotou, private
communication).  The spectra of SGR~1627$-$41 and SGR~1806$-$20 can be modeled
with a power-law (Woods et al.\ 1999a; Mereghetti et al.\ 2000) while
SGR~1900$+$14 requires an additional blackbody component ($kT$ $\sim$0.5 keV
[Woods et al.\ 1999b; Kouveliotou et al.\ 2001]).  The presence of a blackbody
component in SGR~0526$-$66 is marginal (Kulkarni et al.\ 2003).  Currently, we
are investigating the stability of the power-law photon index in
SGR~1900$+$14.  Our preliminary finding is that the photon index evolves from
$\sim$1.0 to $\sim$2.2 within a few years.

\vspace{-0.45in}
\begin{table}[!h]
\begin{minipage}{1.0\textwidth}
\begin{center}
\caption{Quiescent spectral and temporal properties of the Soft Gamma
Repeater X-ray counterparts.  See text for references.}
\vspace{10pt}
\begin{tabular}{lcccc} \hline \hline
      & SGR~0526$-$66 & SGR~1627$-$41 & SGR~1806$-$20 & SGR~1900$+$14 \\\hline
$L_{\rm x}$ (cgs) & 1$\times10^{36}$ & $\lesssim3\times10^{34}$ & 
      4$\times10^{35}$ & 2$\times10^{35}$ \\
$\Gamma$  & 3.1 & 2.5 & 2.0$-$2.2 & 1.0$-$2.2  \\
$kT$ (keV) & 0.5 & ... & ... & ...  \\
$N_{\rm H}$ ($10^{22}$ cm$^{-2}$) & 0.54 & 7.7 & 6.3 & 2.4  \\\hline
Period (s) & 8.0 & ... & 7.5 & 5.2  \\
Pdot (10$^{-11}$ s s$^{-1}$) & 6.5 & ... & 8$-$30 & 6$-$35  \\
\hline\hline
\end{tabular}
\end{center}
\end{minipage}\hfill
\end{table}

Three SGRs show low-amplitude pulsations in their persistent emission within a
narrow period range (Kouveliotou et al.\ 1998a; Hurley et al.\ 1999b; Kulkarni
et al.\ 2003).  The frequency of these pulsations is increasing rapidly
(Kouveliotou et al.\ 1998a, 1999; Kulkarni et al.\ 2003), consistent with the
interpretation of an underlying strongly magnetized neutron star.  Due to the
strong timing noise present in both SGR~1806$-$20 and SGR~1900$+$14 (Woods et
al.\ 2002), it is best to give ranges for their measured period derivatives
(Table 1).

\section{ BURST-INDUCED EFFECTS ON THE PERSISTENT EMISSION}

During the last few years, we have noted changes in the X-ray emission
properties of SGRs during episodes of burst activity (Woods et al.\ 2001). 
Through studying the transient effects imparted upon SGRs (or the lack thereof)
during times of burst activity, we have gained deeper insight into the nature
of the burst mechanism and the SGR systems in general.  The changes in SGR
pulsed properties and persistent X-ray emission and their relationship to burst
activity are presented below.

\subsection{Torque Changes}

Coherent pulsations from the persistent emission of SGR~1806$-$20 were
discovered within an {\it RXTE} PCA observation from 1996 November (Kouveliotou
et al.\ 1998a).  Including both archival and subsequent monitoring observations,
the spin frequency history of SGR~1806$-$20 has now been extended from 1993
through 2001 (Figure 3, from Woods et al.\ 2002).  We have found that at all
times, the SGR has been spinning down, but the rate of spindown shows
substantial variability.  In fact, the measured spin-down torque on this SGR
has been found to vary by up to a factor $\sim$6.  The torque variations seen
in SGR~1806$-$20 do not correlate directly with the burst activity (Woods et
al.\ 2002).

\vspace{-0.4in}

\begin{figure}[!ht]
\centerline{
\psfig{file=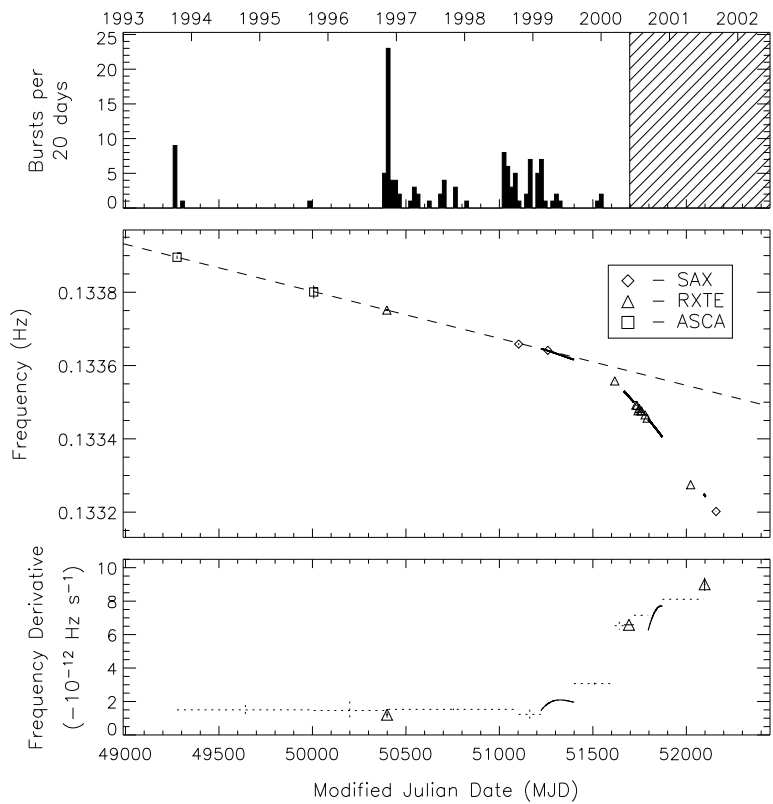,height=4.6in}}
\vspace{-0.33in}

\caption{{\it Top} -- Burst rate history of SGR~1806$-$20 as observed with
BATSE.  The hashed region starts at the end of the {\it CGRO} mission.  {\it
Middle} -- The frequency history of SGR~1806$-$20 covering 8 years.  Plotting
symbols mark individual frequency measurements and solid lines denote
phase-connected timing solutions.  The dashed line marks the average spin-down
rate prior to burst activation in 1998.  {\it Bottom} -- The frequency
derivative history over the same timespan.  Dotted lines denote average
frequency derivative levels between widely spaced frequency measurements. 
Solid lines mark phase-coherent timing solutions and triangles mark
instantaneous torque measurements, both using {\it RXTE} PCA data.}

\end{figure}

Pulsations from the X-ray counterpart of SGR~1900$+$14 were discovered during
an ASCA observation in 1998 April (Hurley et al.\ 1999b), shortly before the
SGR entered an intense, sustained burst active interval (Figure 1).  Similar to
SGR~1806$-$20, a compilation of data preceeding and following the discovery
observation showed that this SGR was spinning down rapidly and irregularly
(Kouveliotou et al.\ 1999; Woods et al.\ 2002) .  As with SGR~1806$-$20, the
variations in torque do not directly correlate with the burst activity from
this SGR with one notable exception, the giant flare of August 27 (Woods et
al.\ 1999c; Palmer 2002).

We note that although a rapid spin-down event most likely occurred within the
hours following the August 27 flare, its impact on the overall spin history of
the SGR was very small relative to the much larger variations observed during
$\sim$5 years of monitoring.  So, in general, the direct effects of burst
activity are insignificant to the overall torque noise in each of these SGRs. 
For a more complete discussion of the torque variability in these two SGRs, as
well as a quantitative analysis of the torque noise, see Woods et al.\ (2002).

Recently, the 8 sec pulsations observed in the tail of the March 5$^{\rm th}$
flare from SGR~0526$-$66 (Mazets et al.\ 1979) were confirmed in the persistent
X-ray emission of the source (Kulkarni et al.\ 2003).  As of yet, there are
only two period measurements which provides a measure of the spindown.  Further
data are required to determine if the torque varies in this system as well.

\subsection{Pulse Profiles}

Currently, the pulse profiles of both SGR~1900$+$14 and SGR~1806$-$20 are very
nearly sinusoidal (i.e.\ they show very little power at the higher harmonics). 
This has not always been the case, however, as both SGRs have shown significant
changes in their pulse profiles during the last several years.  The most
notable of these changes occurred in the pulse profile of SGR~1900$+$14 during
the tail of the giant flare of August 27 (e.g., Feroci et al.\ 2001).

\vspace{-0.7in}

\begin{figure}[!ht]

\centerline{
\psfig{file=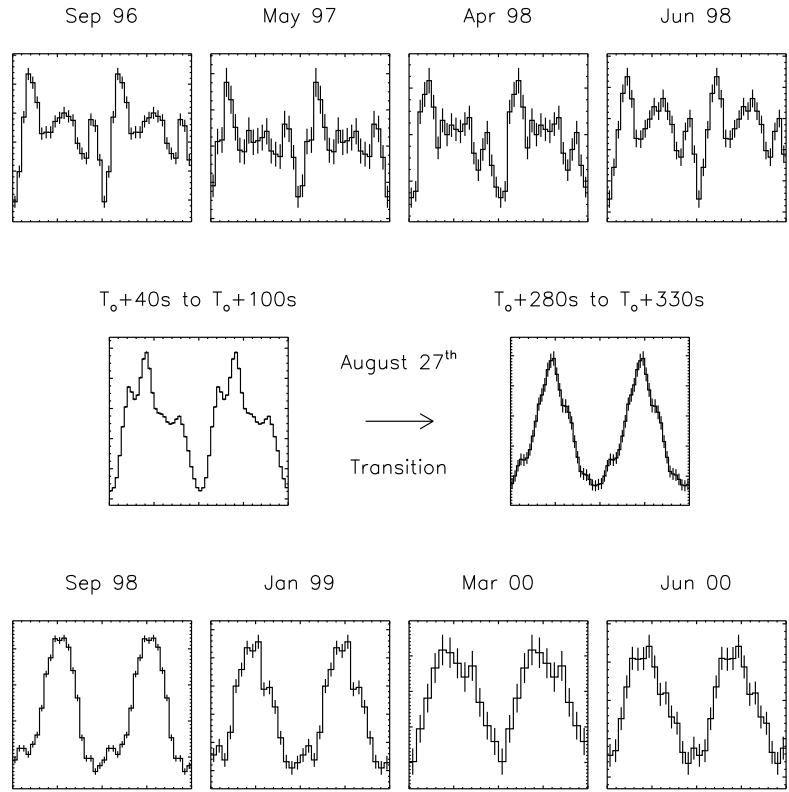,height=5.5in}}
\vspace{-1.0in}

\caption{Evolution of the pulse profile of SGR~1900$+$14 over the last 3.8
years.  All panels display two pulse cycles and the vertical axes are count
rates with arbitrary units.  The two middle panels were selected from Ulysses
data (25$-$150 keV) of the August 27$^{\rm th}$ flare.  Times over which the
Ulysses data were folded are given relative to the onset of the flare (T$_{\rm
o}$).  See text for further details.  The top and bottom rows are integrated
over the energy range 2$-$10 keV.  From top-to-bottom, left-to-right, the data
were recorded with the RXTE, BeppoSAX, ASCA, RXTE, RXTE, RXTE, BeppoSAX, and
RXTE.}

\end{figure}

Forty seconds after the onset of the August 27 flare, 5.16 s coherent gamma-ray
pulsations at high amplitude emerged.  Initially, the pulse profile was
complex, having four distinct maxima per rotation cycle.  A few minutes later
toward the end of the flare, the pulse profile was significantly more
sinusoidal (Figure 4 -- middle row).  The same qualitative behavior was
observed in the {\it persistent} X-ray pulse profile of SGR~1900$+$14.  In all
observations prior to the August 27$^{\rm th}$ flare, the pulse profile was
complex having significant power at higher harmonics (Figure 4 -- top row). 
For all observations after 1998 August 27 through the present, the pulse
profile remained relatively simple (Figure 4 -- bottom row).  Hence, the pulse
profile change observed at gamma-ray energies during the tail of the August
27$^{\rm th}$ flare translated to the persistent emission pulse profile of this
SGR in a sustained manner (i.e.\ for years after the August 27 X-ray tail had
disappeared [Woods et al. 2001]). 

A systematic study of the temporal and spectral evolution of the pulse profiles
of SGR~1900$+$14 and SGR~1806$-$20 is presented in {G\"o\u{g}\"u\c{s}} et al.\
(2002).  This study confirms the temporal evolution of the pulse profile of
SGR~1900$+$14 mentioned above and shows that there is a similar time-dependence
in the pulse profile of SGR~1806$-$20.  From 1996 November to 1999 January, the
pulse profile of SGR~1806$-$20 becomes more sinusoidal.  Due to the sparseness
of the observations, however, we cannot determine the exact time of this
change, nor the timescale over which it progressed to better than 2.3 years.

\vspace{-0.4in}

\begin{figure}[!ht]

\centerline{
\psfig{file=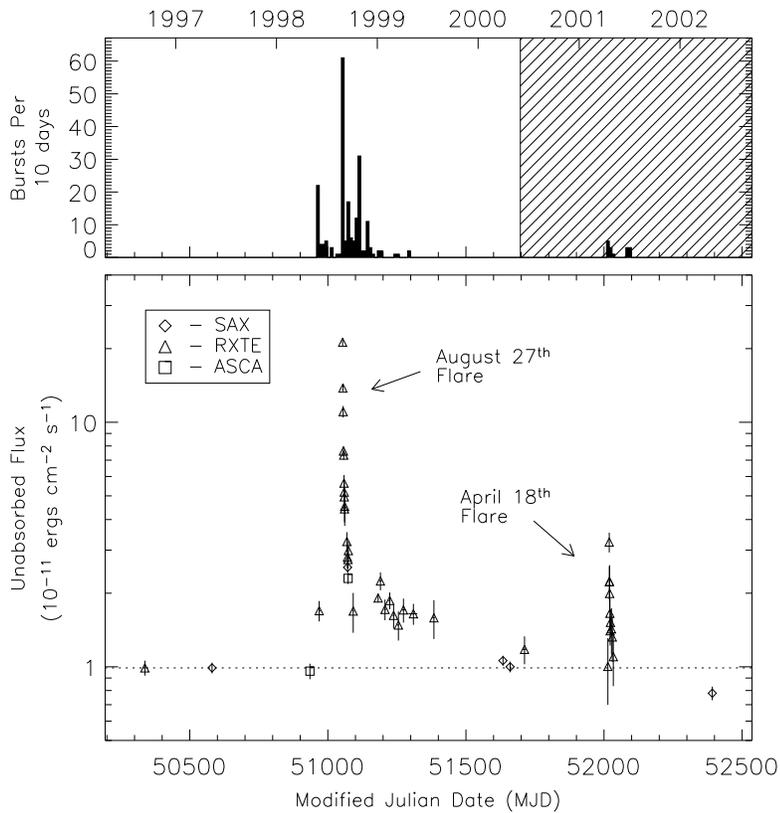,height=4.6in}}
\vspace{-0.35in}

\caption{{\it Top panel} -- Burst rate history of SGR~1900$+$14 as observed
with BATSE.  {\it Bottom panel} -- Persistent/Pulsed flux history of
SGR~1900$+$14 covering 5.5 years.  The vertical scale is unabsorbed 2$-$10 keV
flux.  The pulsed fraction is assumed constant to convert pulsed flux to
phase-averaged flux (see text for details). The dotted line marks the nominal
quiescent flux level of this SGR.  Note the drop in flux below the quiescent
level during the latest observation in 2002.}

\end{figure}

\subsection{Flux Changes}

Changes in the flux of SGRs were first noted in SGR~1900$+$14 following the
giant flare of August 27 (Remillard et al.\ 1998; Kouveliotou et al.\ 1999;
Murakami et al.\ 1999; Woods et al.\ 1999b).  Following this discovery, a
compilation of persistent and pulsed flux measurements over several years
(Figure 5 [Woods et al.\ 2001]) revealed that, in general, there is an
excellent correlation between burst activity (top) and enhancements in the
persistent/pulsed flux from this SGR (middle).  We have found that the pulse
fraction is consistent with remaining constant at most epochs despite changes
in the persistent flux.  It is by assuming that this fraction remains constant
at all times that we can plot both the pulsed flux ({\it RXTE} PCA) and the
persistent flux ({\it BeppoSAX} and {\it ASCA}) on the same scale.  We note,
however, that there are exceptions to this rule when the pulse fraction has
increased for short periods of time (see below).

\subsubsection{\underline{Four Bursts from SGR~1900$+$14}}

The brightest pulsed/persistent flux excess seen in Figure 5 is directly linked
with the August 27 flare.  The flux decays approximately as a power-law in time
($F \propto t^{-0.7}$) following the giant flare (Figure 6 [Woods et al.\
2001]).  The X-ray spectrum as measured with the PCA one day after the flare
can be fit with the sum of a blackbody and a power-law.  The blackbody
temperature is hotter ($kT = 0.94\pm0.03$) than the quiescent level and the
power-law photon index is steeper ($\Gamma = 2.76\pm0.07$).  The spectrum
(0.1$-$10 keV) of the X-ray tail at $\sim$19 days after the flare was found to
be exclusively non-thermal (Woods et al.\ 1999b).  These observations are
consistent with the thermal excess fading more rapidly to its quiescent value
than the power-law component.

\vspace{-0.3in}

\begin{figure}[!ht]

\centerline{
\psfig{file=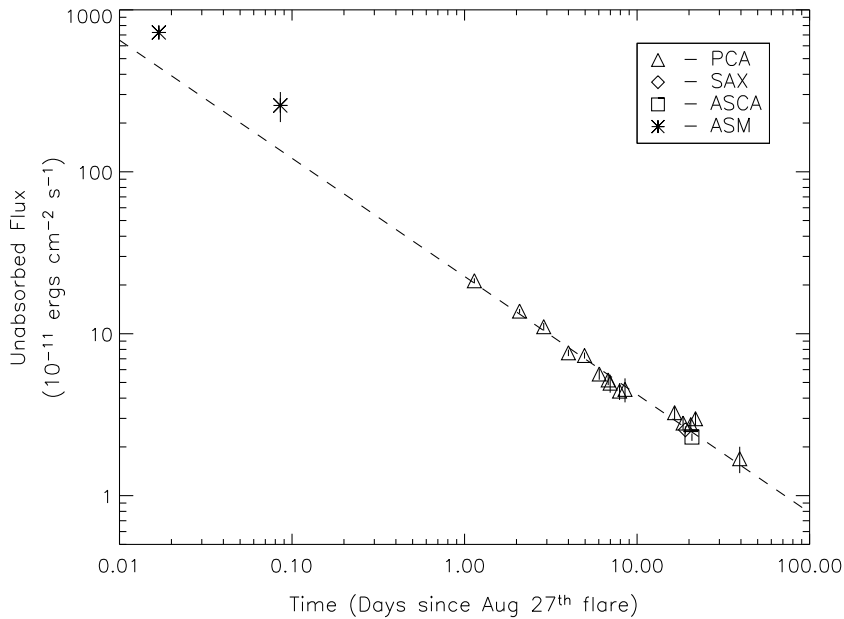,height=3.5in}}
\vspace{-0.15in}

\caption{The flux decay following the 1998 August 27 flare from SGR~1900$+$14. 
The reference time is the beginning of the flare as observed in soft
$\gamma$-rays.  The dotted line is a fit to the RXTE/PCA, BeppoSAX, and ASCA
data only (i.e.\ the ASM data are not included in the fit).  The slope of this
line is $-$0.713 $\pm$ 0.025.}

\end{figure}

For SGR~1900$+$14, there are now four X-ray tails that can be linked with
specific bursts or flares.  The second of these events was recorded on 1998
August 29 (Ibrahim et al.\ 2001; Lenters et al.\ 2003), just two days after the
flare.  This burst had a high gamma-ray fluence and an X-ray tail whose flux
decayed more rapidly than a power-law in time (Lenters et al.\ 2003).  We
define the onset of the tail emission when the intensity quickly drops by
$\sim$4 orders of magnitude within a few tens of milliseconds.  Simultaneously,
the X-ray spectrum suddenly hardens.  Over the next several minutes as the flux
decays through the tail, the energy spectrum softens.  Formally, the energy
spectrum of the tail is equally well fit by a power-law plus a blackbody or an
optically thin thermal bremsstrahlung, each with interstellar attenuation
(Ibrahim et al.\ 2001).  However, the bremsstralung model yields a column
density $\sim$5 times larger than the measured column from the persistent
emission whereas the two component model fit yields a column consistent with
the persistent emission value.  The pulsed fraction increases above the
quiescent level (11\% RMS) up to $\sim$20\% during this tail (Lenters et al.\
2003), and the phase of the pulsations does not shift during the tail relative
to the pre-burst pulse phase.

On 2001 April 18, a burst was detected from SGR~1900$+$14 with a high fluence
($\sim$10$^{43}$ ergs) and long duration $\sim$40 s (Guidorzi et al.\ 2001). 
The burst energy of April 18 was intermediate between the giant flare of 1998
August 27 and typical SGR events.  Like the August 27 flare, this event showed
an extended X-ray tail following the burst which lasted for several days
(Kouveliotou et al.\ 2001; Feroci et al.\ 2003).  The properties of this burst
tail (Table 2) are similar in many respects to the two previously mentioned. 

A second burst from SGR~1900$+$14 also recorded during the 2001 April
activation also possessed an X-ray tail.  This event was detected on April 28
and like previous tails, showed a transient increase in pulse fraction and a
cooling blackbody spectral component (Lenters et al.\ 2003).  In fact, this is
the only tail thus far to show an exclusively thermal X-ray tail.

\vspace{-0.45in}
\begin{table}[!h]
\begin{minipage}{1.0\textwidth}
\begin{center}
\caption{Spectral and pulsed properties of the four X-ray tails
following bright bursts from SGR~1900$+$14.  The burst energy of the events
decreases from left to right.  See text for references.}
\vspace{10pt}
\begin{tabular}{lcccc} \hline \hline
      & 980827 & 020418 & 980829 & 020428 \\\hline
X-ray & BB$+$PL early & BB$+$PL early & BB$+$PL & BB only  \\
Spectrum & PL late & PL late &    &     \\\hline
$kT$ change & Increase & Increase & Increase & Increase  \\\hline
$\Gamma$ change & Steeper & Nothing  & Steeper & N/A  \\
                &         & measured &         &      \\\hline
$E_{\rm tail}/E_{\rm burst}$ & 0.021 & 0.021 & 0.024 & 0.024  \\\hline
Pulse    & Dramatic & Slight & Early   & Nothing   \\
Profile  & change   & change & changes & measured  \\\hline
Pulse    & No change     & Increase to  & Increase to  & Increase to   \\
Fraction & at late times & $\sim$18\%   & $\sim$20\%   & $\sim$32\%    \\\hline
Pulse    & Sudden   & Some    & Nothing  & Nothing   \\
Timing   & spindown & anomaly & measured & measured  \\\hline
\hline
\end{tabular}
\end{center}

\end{minipage}
\end{table}

Within the group of four X-ray tails following bright bursts detected from
SGR~1900$+$14 (Table 2), we find a number of similarities and some slight
differences indicating possible trends (within the restrictions of small number
statistics).  In all cases, the thermal component becomes hotter while the
power-law component brightens following three of the four bursts.  The ratio of
tail energy to burst energy is consistent with being constant among the group
(Figure 7 [Lenters et al.\ 2003]).  Timing anomalies and pulse profile changes
are seen following the two most energetic bursts and the pulse fraction
increases for the three weakest.  An interesting trend which arises from this
small set of X-ray tails is that the pulse fraction enhancement in the separate
X-ray tails appears to correlate with the magnitude of the thermal contribution
to the X-ray flux.  That is, tails with the highest relative blackbody flux
show the largest increase in pulse fraction.  Given the small numbers, however,
more examples are required to establish this trend.

\vspace{-0.35in}

\begin{figure}[!htb]

\centerline{
\psfig{file=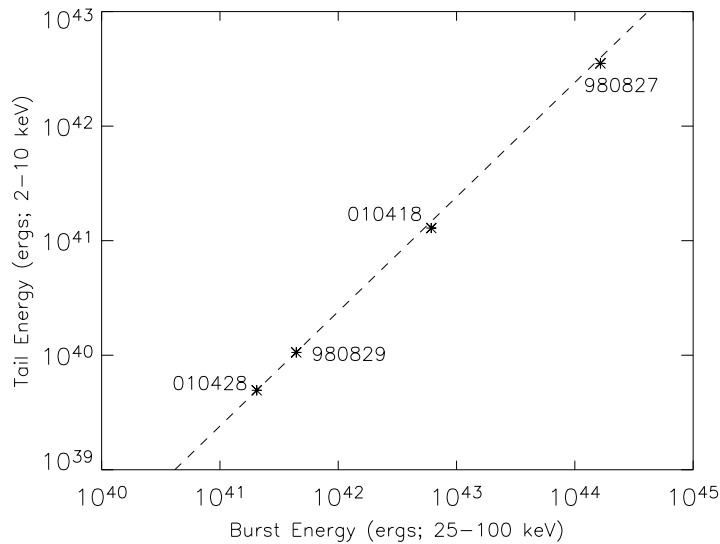,height=3.0in}}
\vspace{-0.2in}

\caption{Tail energy versus burst energy output of four separate
bursts from SGR~1900$+$14.  An assumed distance of 14 kpc is used.}

\end{figure}

\subsubsection{\underline{SGR~1627$-$41}}

SGR~1627$-$41 was discovered in 1998 June when it was observed to burst more
than 100 times in a short span of about a month (Kouveliotou et al.\ 1998b;
Woods et al.\ 1999a).  No bursts from this SGR have been seen since 1998
August.  Shortly after the onset of burst activity in this SGR, an X-ray
counterpart was identified and observed to fade in brightness over the next two
years.  Dissimilar to the observed flux variations in SGR~1900$+$14, the decay
in the lightcurve of SGR~1627$-$41 proceeds on a much longer timescale of order
years and has yet to ``bottom out'' at a baseline flux level.  This object
shows that the response of SGRs to burst activity varies between sources.

\section{ DISCUSSION}

We have summarized the recent observations of dynamic behavior in the
persistent and pulsed emission from SGR~1900$+$14, SGR~1806$-$20 and
SGR~1627$-$41.  Now, we will discuss what constraints these observations place
on the models for the SGRs, in particular the magnetar model.

The torque enhancements in SGR~1806$-$20 and SGR~1900$+$14 do not directly
correlate with the burst activity.  In the context of the magnetar model, the
abscence of a direct correlation between these two parameters has strong
implications for the underlying physics behind each phenomenon.  The magnetar
model postulates that the bursting activity in SGRs is a result of fracturing
of the outer crust of a highly magnetized neutron star.  Furthermore, the
majority of models proposed to explain the torque variability in magnetars
invoke crustal motion and/or low-level seismic activity (Thompson and Blaes
1998; Harding et al.\ 1999; Thompson et al.\ 2000).  Since there is no direct
correlation between the burst activity and torque variability, then either
($i$) the seismic activities leading to each observable are decoupled from one
another, or ($ii$) at least one of these phenomena is {\it not} related to
seismic activity (Woods et al.\ 2002).

The dramatic change in the pulse profile of SGR~1900$+$14 in conjunction with
the giant flare requires a substantial change in the magnetic field of the
neutron star (Woods et al.\ 2001; Thompson et al.\ 2002).  In the magnetar
model, there are at least two possible ways that this can happen.  One
possibility proposed by Thompson et al.\ (2002) is that a twist in the
magnetosphere is generated following the flare, driving a persistent current
which produces an optically thick scattering screen at some substantial
distance ($\sim$10 $R_*$) from the stellar surface.  In this model, the surface
field geometry remains complex at all times.  The pulse profile, however,
simplifies when the scattering screen is present (i.e.\ after the flare).  The
scattering screen must have the properties of redistributing the radiation in
phase, but not in energy in order to account for the reemergence of the
blackbody component after the August 27 tail fades away (Thompson et al.\
2002).  The decay of this magnetospheric twist is believed to be several
years.  An alternative scenario involves restructuring of the surface magnetic
field geometry.  In this picture, the field geometry is complex prior to the
flare and relaxes to a more dipolar structure following the event giving rise
to the observed change in pulse profile.

Thompson et al.\ (2002) recently investigated each of these scenarios in
detail, noting advantages and disadvantages for each model.  In this work, they
have identified further observational tests involving the energy spectrum of
the emission before and after the flare.  Simulations of the expected behavior
and an analysis of the spectral evolution of SGR~1900$+$14 are currently
underway.

Currently, only four clear X-ray tails have been detected from SGR~1900$+$14. 
As mentioned earlier, there is a potential correlation between the relative
abundance of thermal emission in the tail and the enhancement of the pulse
fraction.  This correlation, if proven correct with the detection and analysis
of several more SGR tails, would provide a strong argument for heating of a
localized region on the neutron star during bursts (Thompson et al.\ 2000;
Lyubarsky et al.\ 2002).

Since the pulse fraction increases during three of the four X-ray tails
detected from SGR~1900$+$14, the flux enhancement must be anisotropic about the
star in these cases.  For the August 29 and April 28 burst tails, the location
of the heating is also constrained (Lenters et al.\ 2003).  In each of these
events, we have precise pulse phase information prior to, during, and after the
tail.  For both bursts, the phase of the pulsations during the tail does not
shift relative to the pulse phase prior to the burst.  This requires that the
localized region on the neutron star with the largest relative flux enhancement
is within or nearby the region giving rise to the persistent X-ray pulse peak
(e.g.\ the polar cap).  In the magnetar model, the burst emission is due to the
build up of stress in the stellar crust from the evolving magnetic field and
the eventual release of this stress when the crust fractures (Thompson and
Duncan 1995).  Localized heating of the polar cap requires that the fracture
site of the burst be at the same location.  Since these two events do not occur
at all near the same pulse phase (Lenters et al.\ 2003), the burst emission
arising from near the surface must be significantly scattered.  An alternative
model (Lyutikov 2002) invokes a magnetospheric rather than crustal origin for
the bursts.  This model is attractive in that it negates the need for large
angle scattering of the burst emission and provides a natural explanation for
the preferential brightening of the polar cap regions.

Contrary to the short-lived X-ray tails seen following SGR~1900$+$14 bursts,
the much longer timescale flux decay in SGR~1627$-$41 following its 1998
outburst shows that the effects of burst activity on the persistent emission of
SGRs is not uniform among the class.  As another example, the recent outburst
of the AXP 1E~2259$+$586 (Kaspi et al.\ 2002) shows several source parameters
that changed as a result.  Some of the observed changes (e.g.\ persistent and
pulsed X-ray flux increases) are similar to those seen in SGR~1900$+$14 whereas
others are not.  As discussed above, episodes where the properties of these
objects are rapidly evolving are paramount in determining the nature of these
sources and the physical mechanism driving the dynamics of the source spectral
and temporal properties.  The diversity in SGR/AXP outbursts observed thus far
necessitates detailed study of each outburst in order to obtain a comprehensive
picture of the class as a whole.  This in turn requires diligent monitoring and
prompt follow-up of each source when they become active.

\section{Acknowledgements} I thank the many collaborators who have contributed
to the results discussed here.  I acknowledge financial support from the Long
Term Space Astrophysics program (NAG 5-9350).  I would like to thank Chryssa
Kouveliotou for useful discussions and a careful reading of the manuscript.



\vspace{0.3truein}

\noindent{E-mail address of P.M.\ Woods \hspace{11pt} 
peter.woods@nsstc.nasa.gov}

\end{document}